\documentclass[aps,prl,twocolumn,groupedaddress]{revtex4}
\usepackage{amsmath}
\usepackage{amssymb}
\usepackage{graphicx}

\begin{document}

\title{Anisotropic photoconductivity in graphene}

\author{Maxim Trushin and John Schliemann}

\affiliation{Institute for Theoretical Physics, University of Regensburg,
D-93040 Regensburg, Germany}

\begin{abstract}
 We investigate the photoconductivity of graphene within 
the relaxation time approximation. In presence of the inter-band transitions 
induced by the linearly polarized light
the photoconductivity turns out to be highly anisotropic due to 
the pseudospin selection rule for Dirac-like carriers.
The effect can be observed in clean undoped graphene samples and 
be utilized for light polarization detection.
\end{abstract}

\maketitle

\section{Introduction}

\begin{figure}
\includegraphics[width=\columnwidth]{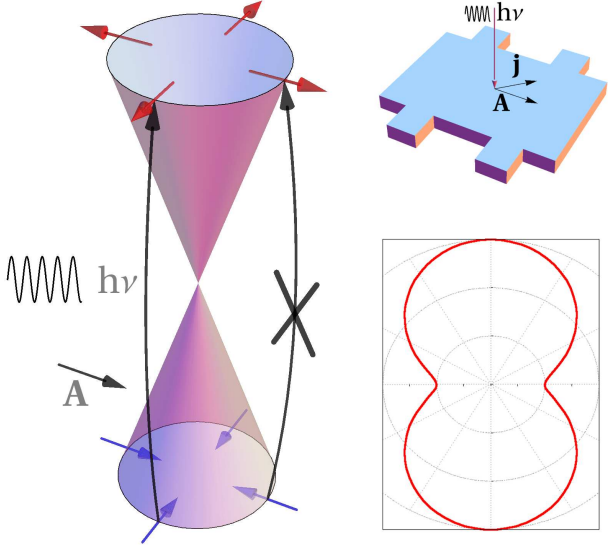}
\caption{Graphene Hall bar sample irradiated by linearly polarized 
electromagnetic wave described by vector potential $\mathbf{A}$.
Applying a bias voltage leads to an electrical current 
$\mathbf{j}$
which depends on the photo-induced carrier concentration.
The pseudospin orientation of the charge carriers described by Dirac 
Hamiltonian with the cone-shaped dispersion law shown by arrows is 
entangled with the particle momentum.
The electrons in the valence band absorbing the photon energy $h\nu$
are excited to the conduction band producing the photoconductivity response.
The electron-hole excitation rate is zero if
the light is polarized along the pseudospins of the excited particles.
In contrast, the excitation rate is maximal if the vector
potential and pseudospin are perpendicular to each other.
Since the pseudospin orientation is coupled with the particle's momentum
the resulting photoconductivity $\sigma_\mathrm{ph}$ depends on the angle between $\mathbf{A}$
and $\mathbf{j}$ as shown in the inset.
The absolute value of $\sigma_\mathrm{ph}$ is estimated in fig.~\ref{fig2}.}
\label{fig1}
\end{figure}

Graphene membranes are optically transparent\cite{Science2008nair} as well as 
highly conductive\cite{SSC2008bolotin}
even at room temperatures\cite{PRL2008bolotin}.
These two properties being incompatible with each other
in conventional materials occur in carbon monolayers
quite naturally and make them very promising for
optoelectronical applications.\cite{NL2010avouris,Natphot2010review}
There is, however, another unusual property of carriers in graphene 
which makes this material even more interesting for optoelectronics.
The carriers in graphene display an additional degree of freedom
which is often dubbed as the pseudospin but, in fact, is connected to the 
sublattice index and has nothing
to do with the real spin.\cite{RMP2009castroneto}
We show, that the pseudospin manifests itself in the inter-band
optical absorption making the transition probability sensitive
to the pseudospin orientations in the initial and final states
in a way similar to the real spin selective rules for the inter-band optical 
transitions in III-V semiconductors.
Since the pseudospin is textured in the momentum space, as shown in 
fig.~\ref{fig1}, graphene's photoconductivity
turns out to be anisotropic in the case of the linearly polarised light.
The effect seems to be strong enough to find some applications in graphene 
optoelectronics.

The model described below involves the optical excitation of the valence 
electrons to the conduction band of intrinsic (i.e. undoped) graphene.
The idea is that the effective Hamiltonian describing the interaction between 
the electromagnetic wave and
carriers in graphene inherits the pseudospin-momentum entangled structure 
from the low energy kinetic term derived
within the tight-binding approach.\cite{RMP2009castroneto}
Assuming normal incidence of a linear polarized electromagnetic wave one 
deduces an electron generation rate
which strongly depends on the relative orientation between the electron 
momentum and the linear polarization plane, see fig.~\ref{fig1}.
As consequence, the photoconductivity is predicted to be anisotropic
resulting in 
a high on/off ratio as a function of the linear polarization angle.
We note that the photoconductivity in graphene has been also theoretically 
investigated in recent works,\cite{PRB2008vasko,PRB2010romanets}
not analyzing its anisotropy.
Moreover, the photoconductivity studied in this work should not be confused 
with the photocurrents predicted  \cite{PRB2008syzranov,PRB2009oka,PRB2010entin,PRB2011mai}
and measured \cite{Natnano2009xia,NL2010xu,NL2009park,PRL2010karch} in graphene.
The photocurrent can be generated without bias voltage applied, whereas
the bias is necessary for the photoconductivity measurements.
The photoconductivity and photocurrent anisotropy has been also found
in the materials\cite{JETPL1985esayan,PTP1982karaman,JETP1969galperin} other than graphene.

\section{Preliminaries}
The two-band effective Hamiltonian for $\pi$-system of graphene near 
half filling
is $H_0=\hbar v_F (\sigma_x k_x+\sigma_y k_y)$,
where $v_F\approx 10^6 \mathrm{ms^{-1}}$, ${\mathbf k}$ is 
the electron momentum, and
$\sigma_{x,y}$ are the Pauli matrices.
The Pauli operator $\vec{\sigma}$
represents the pseudospin orientation which is depicted in fig.~\ref{fig1}
for the eigenstates of $H_0$ given by
$\Psi_{\kappa\mathbf{k}}(x,y)=\frac{1}{\sqrt{2}}{\mathrm e}^{ik_x x+ik_y y}
\left(1,\kappa {\mathrm e}^{i\theta}\right)^T$,
where $\tan\theta=k_y/k_x$, and $\kappa=\pm$ denotes the band index, and
the energy spectrum of $H_0$ is $E_{\kappa k}=\kappa \hbar v_F k$.

The interaction  between the electromagnetic wave 
and charge carriers is described by the Hamiltonian
$H_\mathrm{int}= \frac{e v_F}{c}(\sigma_x A_x+\sigma_y A_y)$
and resembles the pseudospin structure.
Assuming the {\em normal} incidence and {\em linear polarization} of the 
electromagnetic wave
$\mathbf{A}=\mathbf{A}_0 \exp(-i\omega t+ ik_z z)$ the golden-rule
inter-band transition rate reads
\begin{equation}
\label{Ik}
I[f_{\kappa \mathbf{k}}]=
\sum_{\kappa'} \int\frac{d^2 k' L^2}{4\pi^2}
w(\kappa' \mathbf{k}',\kappa \mathbf{k})(f_{\kappa' k'}-f_{\kappa k}),
\end{equation}
where  $f_{\kappa k}$ is the distribution function, and
\begin{eqnarray}
\nonumber && w(\kappa' \mathbf{k}',\kappa \mathbf{k})
=\frac{2\pi}{\hbar}\frac{4\pi^2}{L^2}
\delta(k_x-k'_x)\delta(k_y-k'_y) \left(\frac{ev_F}{c}\vert A\vert \right)^2 \\
\nonumber
&&\times \left[\delta(E_{\kappa' k'} - E_{\kappa k} -\hbar\omega) 
+ \delta(E_{\kappa' k'} - E_{\kappa k} +\hbar\omega) \right]\\
&&\times \frac{1+\kappa\kappa'\cos(\theta+\theta'-2\theta_\mathrm{pol})}{2}
\label{probabil}
\end{eqnarray}
is the transition probability. Here $\omega=2\pi\nu$ is the radiation frequency,
and $\tan \theta_\mathrm{pol}=A_y/A_x$ is the linear polarization angle.
The length $L$ plays a role of the sample size or the laser spot diameter
whichever is smaller.
Eq.~(\ref{probabil}) describes the direct inter-band transitions and, thanks 
to the momentum and energy conservation,
naturally includes $\delta$-functions in the first two lines.
Most important, however, is the third line which depends on the difference
between the linear polarization angle $\theta_\mathrm{pol}$ and direction of carrier motion.
This dependency disappears in the case of the circular polarization and is crucial for the effect considered below.


\section{Photoconductivity within the relaxation time approximation}
In the following we focus on the electron transport, i.e. $\kappa=+$, and
the carriers are excited from the valence to conduction band, as shown in 
fig.~\ref{fig1}.
To describe the recombination process we introduce the inelastic relaxation 
time $\tau_i$ which
corresponds to the life time of the optically excited states.
The steady state distribution function $f^{(1)}_{+k}$
is then obtained by balancing the generation rate
(\ref{Ik}) and the relaxation rate $f^{(1)}_{+ k}/\tau_i$ and reads
\begin{eqnarray}
\nonumber && f^{(1)}_{+k}=\frac{2\pi \tau_i}{\hbar} 
\left(\frac{ev_F}{c}\vert A\vert \right)^2 \delta(E_{- k} - E_{+ k} +\hbar\omega)\\
&& \times \left[f^{(0)}_{- k}-f^{(0)}_{+ k}\right] \sin^2(\theta-\theta_\mathrm{pol}).
\label{f1}
\end{eqnarray}
We naturally assume that the initial state is the equilibrium one
described by the Fermi-Dirac distribution function $f^{(0)}_{\pm k}$.
There is no electrical current in the steady state described
by the distribution function (\ref{f1}).

The momentum relaxation is assumed to be
due to the elastic scattering of carriers on impurities.
The average momentum $\hbar\Delta {\mathbf k}$
which the electrons gain due to the external electric field ${\mathbf E}$
can be estimated as $\hbar \Delta {\mathbf k}  = e  {\mathbf E} \tau_e$,
where $\tau_e$ is the elastic momentum relaxation time.
For small electric field  (linear response)
the non-equilibrium term $f^{(2)}_{+k}$ can be obtained by
expanding the steady-state function $f^{(1)}_{+(k-\Delta k)}$
with respect to small $\Delta\mathbf{k}$ in up to 
linear order in $\mathbf{E}$.
Recalling $\hbar \mathbf{v} = -\partial_{\Delta \mathbf{k}}E_{+(\mathbf{k}-\Delta\mathbf{k})}\vert_{\Delta \mathbf{k}=0}$,
the non-equilibrium distribution function for photo-excited 
electrons $f^{(2)}_{+ k}$ can be written as
\begin{equation}
\label{solution}
f^{(2)}_{+k}=-e\mathbf{E}\mathbf{v}\tau_e
\frac{\partial f^{(1)}_{+k}}{\partial E_{+k}}, 
\quad \mathbf{v}=v_F\left(\begin{array}{c}
\cos\theta \\ 
\sin\theta
\end{array} \right).
\end{equation}
Eq.~(\ref{solution}) is valid if and only if $\tau_i\gg\tau_e$,
i. e. optically excited states live much longer than the average
time between two subsequent elastic scattering events.
This is actually the case in graphene.\cite{NL2010avouris,Natphot2010review}

The current density due to the photo-excited electrons can be written as
$\mathbf{j}_\mathrm{ph}=e\int\frac{d^2 k}{4\pi^2} \mathbf{v} f^{(2)}_{+k}$.
This integral can be calculated in polar coordinates with the subsequent
substitution $\varepsilon=E_{+k}$ and reads
\begin{eqnarray}
\nonumber &&
\int d\varepsilon \varepsilon \frac{\partial}{\partial \varepsilon}
\left[\delta(\hbar \omega -2 \varepsilon) 
(f^{(0)}_{-\varepsilon} - f^{(0)}_{+\varepsilon}) \right]\\
&&  = -\frac{1}{2}[f^{(0)}_{-\varepsilon} -
f^{(0)}_{+\varepsilon}]_{\varepsilon=\frac{\hbar\omega}{2}}.
\end{eqnarray}
The photoconductivity for a given valley/spin channel is then given by
\begin{equation}
\label{sigma}
\sigma_\mathrm{ph}=A_\mathrm{ph}\left(\begin{array}{cc}
2-\cos(2\theta_\mathrm{pol}) & -\sin(2\theta_\mathrm{pol})  \\ 
-\sin(2\theta_\mathrm{pol}) & 2+\cos(2\theta_\mathrm{pol})
\end{array}\right)
\end{equation}
with the amplitude $A_\mathrm{ph}$ being
\begin{equation}
\label{sigma0}
A_\mathrm{ph}=\frac{e^2}{16\hbar^3}\tau_e\tau_i 
\left(\frac{ev_F}{c}\vert A\vert \right)^2
(f^{(0)}_{-\varepsilon} -
f^{(0)}_{+\varepsilon})\vert_{\varepsilon=\frac{\hbar\omega}{2}}.
\end{equation}
Rigorous analysis based on the Boltzmann equation written
within the relaxation time approximation suggests
the same expression for $\sigma_\mathrm{ph}$ but both $\tau_i$ and $\tau_e$
must be substituted by the total relaxation time
$\tau^{-1}=\tau_e^{-1}+\tau_i^{-1}+...$.
The effect of anisotropy predicted here does not depend on $\tau$ anyway.
Indeed, diagonalizing the matrix (\ref{sigma}),
the photoconductivity $\sigma_\mathrm{ph}^\parallel=A_\mathrm{ph}$
parallel to the light polarization plane turns out to be $3$ times smaller
than the perpendicular one $\sigma_\mathrm{ph}^\perp=3A_\mathrm{ph}$, 
i.e. the photoconductivity
is highly anisotropic, but the anisotropy itself is independent of $\tau$'s.
Thus, changing the linear polarization angle
from $0$ to $2\pi$ one can observe two minima (and two maxima)
in the current flow, as depicted in the inset of fig.~\ref{fig1}.
These double extrema are a key signature of the effect predicted.

\section{Discussion and conclusion}
Let us discuss the conditions necessary to observe 
the anisotropic
photoconductivity given by eq.~(\ref{sigma}) and shown in fig.~\ref{fig1}.
As it is clear from the analysis given in the previous section
the {\em relative} anisotropy does not depend on the relaxation times because
the relaxation processes reduce the overall photoconductivity,
not only its anisotropic part.
The physical reason why the anisotropy does not vanish due to the momentum relaxation is 
the very fact that the anisotropic non-equilibrium distribution relaxes as fast
as its isotropic contribution does.
We believe therefore that the anisotropy can be detected easily
as long as the photoconductivity response is large itself.


To observe the photoconductivity the chemical potential $\mu$ in graphene should be smaller than 
one-half of
the excitation energy $\hbar\omega/2$ enabling direct excitations
from the valence band.
Assuming THz radiation, as used in the work by Karch {\it et al.}, 
\cite{PRB2010krach}
we arrive at the maximum $\mu$ less than $10\,\mathrm{meV}$.
Thus, the unintentional doping in graphene samples used
before\cite{PRB2010krach} should be reduced by almost of two orders of 
magnitude.
The temperature can also affect the effect even if
the sample is perfectly neutral by reducing the photoconductivity 
by a factor of the order of $\hbar\omega/2T$ at
zero chemical potential. Thus, room temperature $T=25\,\mathrm{meV}$
seems to be somewhat to high for observing a sufficient signal at
a radiation frequency of 1 THz.
Moreover, the relaxation times $\tau_e$ and $\tau_i$
assumed to be constant so far,  will in fact also be 
temperature-dependent.
However, one can facilitate the measurement by
increasing the overall multiplier proportional to the radiation power,
possibly by means of a high power pulsed $\mathrm{NH_3}$ 
laser.\cite{ganichev2006}

In contrast to the photocurrents due to photon 
drag\cite{PRB2010entin,PRL2010karch,PRB2010krach}
the above effect is due to the pseudospin-selective 
inter-band transitions.
The momentum transfer from photons to carriers is not important,
and the effect should be observable even at normal incidence of light.
The predicted anisotropy is strongest for linearly polarized light source,
whereas for circular polarization the transition probability (\ref{probabil})
does not depend on the direction of carrier motion, and the 
photoconductivity anisotropy does not occur.
An elliptically polarized light source interpolates between these extreme cases.
Moreover, the vanishing anisotropy in the case of circular polarization 
can be used to separate the effect in question from the other photocurrent 
contributions.\cite{Natnano2009xia,NL2010xu,NL2009park,PRL2010karch,PRB2010krach}

It is also interesting that the  anisotropy predicted
\cite{PRB2011mai} and observed\cite{echtermeyer2011} recently in the photocurrent through graphene pn-junctions
seems to have the same origin as the one predicted here.
There is, however, $\pi/2$ off-set in the photocurrent vs. polarisation angle dependency
as compared with the one shown in fig.~\ref{fig1}.
This is probably because ``the resulting photocurrent comes mainly from electrons moving
nearly parallel to the barrier'' \cite{PRB2011mai},
and in order to maximize the concentration of such electrons
the polarization plane must be set perpendicular to the pn-junction,
i.e. along the photocurrent flow.
One can reproduce this $\pi/2$ off-set also within our model
by taking into account the dependence on ${\mathbf k}$ of the angle $\theta$
appearing in eq.~(3) in the driving term of the Boltzmann equation.

As already stated, the eigenvalues of the photoconductivity tensor are 
predicted to differ by a factor of $3$. 
In order to estimate the overall magnitude of the photoconductivity
compared to other conduction mechanisms, let us
compare the residual
carrier concentration due to the unintentional doping
with the one induced by the inter-band excitation.
The former varies from
$10^{11}\,\mathrm{cm^{-2}}$ for low mobility flakes
on $\mathrm{SiO_2}$ to $10^{8}\,\mathrm{cm^{-2}}$
for suspended samples after annealing.\cite{RMP2010peres}
The latter can be estimated as $n_\mathrm{ph}=\tau_i/(L^2 \tau_\mathrm{ph})$
where $\tau_\mathrm{ph}$ relates to the total photo-excitation rate as
$1/\tau_\mathrm{ph}=\int\frac{d^2 kL^2}{2\pi^2}I[f_{+ \mathbf{k}}]$.
On the other hand $\hbar\omega/\tau_\mathrm{ph}$ can also be seen as the 
radiation energy 
absorption rate
which is nothing else than the absorbed radiation power $W_a$.
Note, that $W_a$ relates to the incident radiation power $W_i$ 
as $W_a/W_i=\pi \alpha$  (where $\alpha=e^2/(\hbar c)$ is the fine structure constant)
for a single layer graphene membrane.\cite{Science2008nair,PRL2008kuzmenko}
Thus, $n_\mathrm{ph}$ at finite temperature $T\neq0$ can be estimated as
\begin{equation}
n_\mathrm{ph} = 0.023\frac{ W_i \tau_i}{L^2 \hbar \omega}(f^{(0)}_{-\varepsilon} -
f^{(0)}_{+\varepsilon})\vert_{\varepsilon=\frac{\hbar\omega}{2}}.
\label{nph}
\end{equation}

\begin{figure}
\includegraphics[width=\columnwidth]{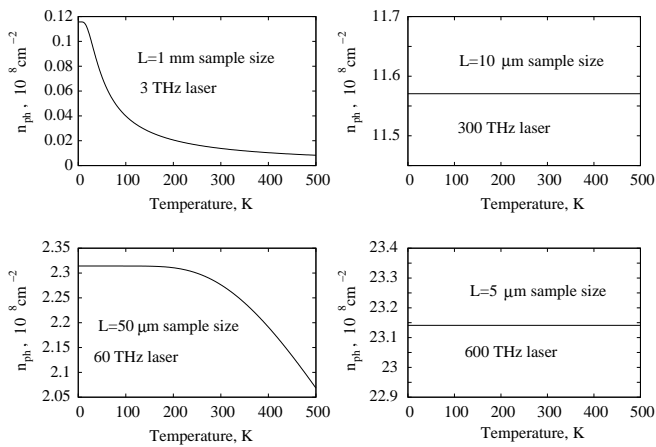}
\caption{The concentration of photoinduced carriers $n_\mathrm{ph}$ as a function of temperature
at different radiation frequency and sample size. (The latter is assumed to be roughly equal to the laser spot diameter
and, therefore, has to be substantially larger than the radiation wavelength to avoid diffraction effects.)
The photoconductivity can be estimated as $\sigma_\mathrm{ph}=e\mu n_\mathrm{ph}$,
where $\mu$ is the mobility of carriers. The incident radiation power $W_i$
and relaxation time $\tau_i$ are $10\,\mathrm{mW}$ and $1\,\mathrm{ps}$ respectively.}
\label{fig2}
\end{figure}

To be specific we assume
that the photoconductivity is generated 
by a $\mathrm{CH_3OH}$ laser\cite{PRB2010krach} with wavelength
$118\,\mathrm{\mu m}$ (i.e. $\hbar\omega=10.5\,\mathrm{meV}$)
and  $W_i\simeq 20\,\mathrm{mW}$,
and the sample itself is
a suspended graphene membrane
of the macroscopic size slightly larger than
the laser spot diameter of about $1\,\mathrm{mm}$.
Assuming 
$\tau_i\simeq 1\,\mathrm{ps}$ \cite{NL2010avouris,Natphot2010review}
we arrive at 
$n_\mathrm{ph}\sim 2\cdot 10^{7}\,\mathrm{cm^{-2}}$ for $L^2\simeq 1\,\mathrm{mm^2}$ and $T=0$.
This values are comparable to the residual carrier concentration
for suspended samples,\cite{RMP2010peres}
thus, the conductivity change in the irradiated graphene
should be observable. Note, that $n_\mathrm{ph}$ can be
substantially increased by utilizing smaller samples
and focusing the laser beam to a smaller spot.
This requires a smaller radiation wave length (i. e. a higher laser frequency)
to avoid diffraction effects. The results are summarized in fig.~\ref{fig2}.

The effect proposed above relies on the pseudospin texture 
shown in fig.~\ref{fig1}.
This texture remains stable as long as the low energy 
one-particle Hamiltonian $H_0$ holds.
At least from a theoretical point of view, the pseudospin texture can
be altered by electron-electron interactions
which may be important in extremely clean samples.\cite{PRL2011trushin}
This is the only fundamental obstacle 
for the photoconductivity anisotropy observation which we can see so far.

To conclude, we predict strong anisotropy of the photoconductivity in graphene
is presence of the linearly polarized light. To observe the effect,
we suggest to use undoped suspended graphene samples
which allow the laser beam to excite the substantial number of photo-carriers
from the valence band.
The cleaner samples are expected to demonstrate the better results.
They can be used as transparent detectors for the polarisation of the
light passing through.

\begin{acknowledgments}

We thank Sergey Ganichev, Vadim Shalygin and Tim Echtermeyer for stimulating discussions.
This work was supported by DFG via GRK 1570 and SFB 689.

\end{acknowledgments}

\bibliography{graphene.bib,optical.bib}

\end{document}